\documentclass[twocolumn,letterpaper,amsmath,amssymb]{revtex4}
\usepackage{graphicx}
\usepackage{wasysym}
\begin{document}

\title{Particle dynamics and effective temperature of jammed granular
\\ matter in a slowly sheared 3D Couette cell}

\author{Ping Wang}
\author{Chaoming Song}
\author{Christopher Briscoe}
\author{Hern\'an A. Makse}

\affiliation {Levich Institute and Physics Department, City College of New York, New York, NY 10031, US}

\date{\today }
\begin{abstract}

We report experimental measurements of particle dynamics on slowly
sheared granular matter in a three-dimensional (3D) Couette cell.
A closely-packed ensemble of transparent spherical beads is
confined by an external pressure and filled with fluid to match
both the density and refractive index of the beads. This allows us
to track tracer particles embedded in the system and obtain
three-dimensional trajectories, $(r(t),\theta(t),z(t))$, as a
function of time. We study the PDF of the vertical and radial
displacements, finding Gaussian and exponential distributions,
respectively.  For slow shear rates, the mean-square fluctuations
in all three directions are found to be dependent only on the
angular displacement of the Couette cell, $\Delta\theta_e$,
$\langle\Delta z^2\rangle\sim\Delta\theta_e$, $\langle\Delta
r^2\rangle\sim\Delta{\theta_e}^\alpha$, $\langle\Delta
\theta^2\rangle\sim\Delta{\theta_e}^\beta$, where $\alpha$ and
$\beta$ are constants. With $\Delta\theta_e$ proportional to the
time between measurements, the values of the constants, $\alpha$
and $\beta$, are found to be sub-diffusive and super-diffusive,
respectively.  The linear relation between $\langle\Delta
z^2\rangle$ and angular displacement implies a diffusive process,
from which we can calculate an ``effective temperature'',
$T_{\mbox{\scriptsize eff}}$, in the vertical direction, through a
Fluctuation-Dissipation relation.  It is of interest to determine
whether these systems can be described by analogous equilibrium
statistical mechanics concepts such as ``effective temperature''
and ``compactivity''. By studying the dynamics of tracer
particles, we find the effective temperature defined by the
Stokes-Einstein relation to be independent of the tracer particle
characteristic features, such as density and size, and dependent
only on the packing density of the system. For slow shear rate,
both the diffusivity and mobility of tracer particles are
proportional to the shear rate, giving rise to a constant
effective temperature, characteristic of the jammed system.
We finally discuss the significance of the existence of
$T_{\mbox{\scriptsize eff}}$ for a statistical mechanics
formulation of granular matter.

\end{abstract}
\maketitle

\clearpage

\section{Introduction}

Fluctuation-Dissipation (FD) relations are commonly used in
equilibrium systems, derived from the notion that small
perturbations and Brownian fluctuations produce the same response
in a given system \cite{landau}. Mobility, the constant of
 proportionality
between a particles drift speed and a constant external force, is
extracted from velocity statistics of particles in a given system.
Diffusivity, calculated from fluctuation displacements of
particles in a system over time, represent the Brownian motion.
The temperature of a system in thermodynamic equilibrium can be
extracted from a FD relation, defined as the ratio of diffusivity
and mobility, as is commonly used in the Einstein relation.  In
equilibrium this temperature is taken to be the bath temperature.

%




As studies in granular matter have grown more important within the
environmental and industrial fields, the need to establish a
scientific framework that accurately predicts granular system
responses on the continuum level, beyond merely geometrical
features, has also escalated.  Granular matter, when condensed to
sufficiently high volume fractions, undergoes a `jamming'
transition to the jammed state. The jammed state is defined as the
condition when a many-body system is blocked in a configuration
far from equilibrium, such that relaxation cannot occur within a
measurable time-scale.  For granular matter, the jammed state
indicates a transition between a solid-like behavior, and a
liquid-like behavior.  At high volume fractions, the physical size
of the constituent grains inhibits particle motion, thereby
rendering the system out of equilibrium, and the granular system
behaves more like a solid.  Thermal motion does not govern the
exploration of states in jammed granular matter.


Theories proposed by Edwards and collaborators \cite{edwards}
propose a statistical mechanics for granular matter based on
jamming the constituent grains at a fixed total volume such that
all microscopic jammed states are equally probable and exhibit
ergodicity.  The exploration of reversible jammed states is
achieved via an external perturbation such as tapping or shear,
not Brownian motion as in thermal systems.  There is an important
difference between reversible jammed states, and states that are
only mechanically stable within certain limits of perturbation
magnitude. For example, pouring grains into a container results
in a pile at a particular angle of repose. This mechanical
equilibrium configuration is jammed regularly but not reversibly jammed because in response to an
external perturbation, the constituent particles will irreversibly
rearrange, approaching a truly jammed configuration.  Studying an
ensemble of truly jammed, reversible states is thereby suitable
for a plausible application of statistical mechanics under the
present theory.  These ensembles, inherently non-equilibrium
systems, will not be governed by the commonly used parameters of
equilibrium statistical mechanics, such as a bath temperature.

In recent studies theoretical mean-field models of glasses
\cite{ckp} have introduced the concept of an ``effective
temperature'' as extracted from the FD relations in
non-equilibrium systems.  While not equivalent to the equilibrium
bath temperature, the effective temperature reflects a change in
the relaxation time-scale of the system.  These non-equilibrium
systems extend beyond glasses, and into granular media, where
physical size of the constituent grains inhibits motion, allowing
for jammed systems far from equilibrium. This concept has been
furthered by computer simulations of granular media and other
non-equilibrium soft-matter systems
\cite{liu3,bklm,sciortino,barrat,ono,mk}.  It remains a
question whether or not granular media can be characterized by an
effective temperature, thus revealing a dynamic counterpart to the
static ``compactivity'' as proposed by Edwards \cite{edwards}.

\begin{figure}
\centering\resizebox{8cm}{!}{\includegraphics{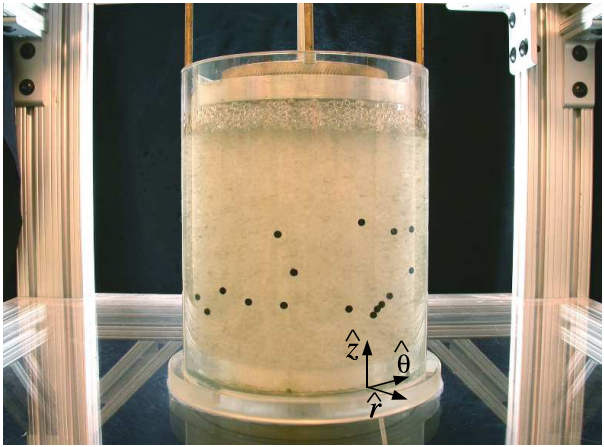}}
\caption{(Color online) Picture of experimental set-up. Transparent acrylic
grains and black tracers in a refractive index and density matched
solution are confined between the inner cylinder of radius
$5.08$cm and the outer cylinder of radius $6.67$cm.} \label{setup1}
\end{figure}

Athermal systems require the input of energy by an external source
to explore the effective temperature \cite{review}. One proposed
method of calculating the effective temperature of a jammed
granular system is a slow shearing procedure
\cite{howell,veje,mueth,mueth2,utter,nedderman,drake}, leading to
the design of the experiment we present in \cite{swm}. Slow
shearing, at the quasi-static limit, allows for extrapolation
towards an effective temperature of jammed, static, systems. The
jammed system of interest is one of identical, spherical grains,
confined between the two cylindrical walls of a 3D Couette cell.
The grains are further confined by an external pressure in the
vertical direction.  The inner cylinder of the Couette cell is
slowly rotated to induce shearing in the system. Tracer particles
are inserted in the system, and their trajectories recorded via
multiple cameras surrounding the system. The Couette cell is
partially filled with a refractive index matched fluid to allow
for system transparency.  The cylindrical walls are roughened by
gluing grains, identical to those of the bulk, such that
crystallization is avoided.

\begin{figure}
\centering\resizebox{8cm}{!}{\includegraphics{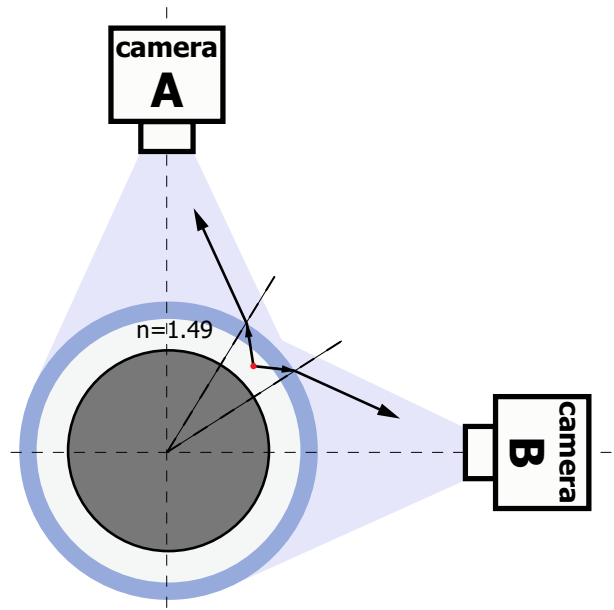}}
\caption{(Color online) Top view
of experimental set-up. The outer cylinder is made of the same
material as acrylic grains ($n\simeq 1.49$). Once the refractive
index is matched, light scattering from tracers will refract only
one times on the outer surface of the outer cylinder. A single particle
is captured by two cameras allowing the determination of the
3-dimensional coordinates of the particle, $(r,\theta,z)$.}
 \label{setup2}
\end{figure}

The cameras record tracer particle trajectories throughout the
bulk, recording data in cylindrical coordinates,
$(r(t),\theta(t),z(t))$.  Distributions of tracer particle
displacements are measured in each direction.  As gravity is the
external force applicable to the mobility calculation in the
current formalism, only displacements in the z-direction are
applied to the FD relation.  Additionally, average velocity
profiles are calculated for each direction, first with constant
shear rate, $\dot{\gamma_e}$, and further studied to determine
shear rate dependence.  Displacement measurements are further
limited to the ``constant mobility and diffusivity'' (CMD) region,
defined as the narrow range of radial coordinates such that the
average vertical velocity is roughly independent on radial
distance.  The PDF of displacement distributions for each
direction is presented.  Further, fluctuations in displacement are
determined for each cylindrical direction, and studied as a
function of time.  Radial displacement fluctuation is found to be
sub-diffusive, while angular displacement fluctuation is found to
be super-diffusive.  Vertical displacement fluctuation is purely
diffusive within the time scales of the experiment, allowing for
the validity of the FD relations used herein.  All displacement
fluctuations are reduced to functions of angular displacement, and
the results are presented.  Such relationships permit scaling of
the PDF curves with varying angular displacements due to changing
shear rates.

Utilizing the FD relations presented above, the diffusivity and
mobility in the z-direction are extracted from the tracer particle
trajectories and the effective temperature is realized. This
effective temperature is found to be independent of tracer
particle properties, as shown in \cite{swm}, and further
independent of the slow-shear rate. Moreover, the effective
temperature may then be considered a physical variable that
characterizes the jammed granular system, with respect to the
generalization of the equilibrium statistical mechanics of Boltzmann,
as applied to non-equilibrium systems.

We further study the limits within which this effective
temperature may be a valid physical variable, as we determine
mobility and diffusivity as a function of shear rate.  While
diffusivity appears independent of shear rate, even somewhat above
the `slow' regime, mobility shows a clear decrease in magnitude as
we explore shear rates above the slow regime, resulting in an
increase in the effective temperature as a function of shear rate.

In this paper, we will further report the experimental detail of
particle dynamics in \cite{swm}.  The outline is as follows:

\section{Experimental Method}
\subsection{Experimental Setup}

\begin{figure}
\centering\resizebox{8cm}{!}{\includegraphics{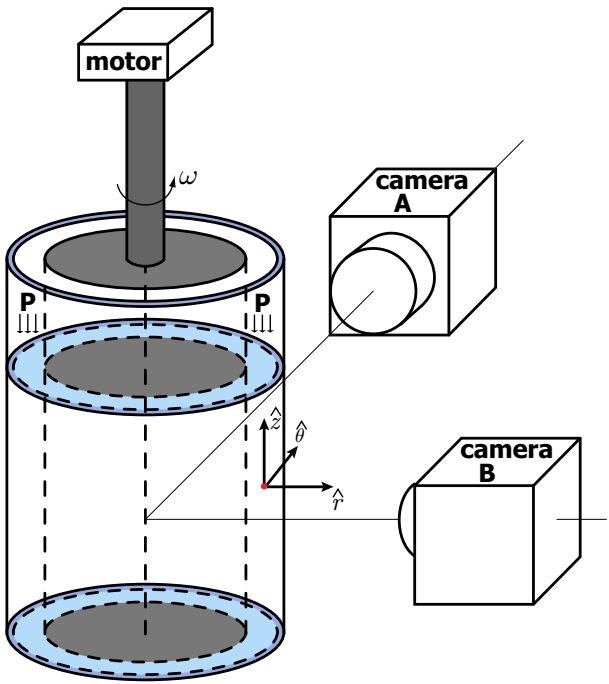}}
\caption{(Color online) Sketch
of experimental set-up. Note that the cylinder is surrounded by 4
cameras, in the sketch we plot only two cameras. A single particle
is captured by two cameras allowing the determination of the
3-dimensional coordinates of the particle, $(r,\theta,z)$.}
 \label{setup3}
\end{figure}

The experiment is performed using a three-dimensional (3D) Couette
cell, as shown in Fig. \ref{setup1}, \ref{setup2} and \ref{setup3}.
The grains are confined between two cylinders of height $19.0$cm.
The inner cylinder is rotated via a motor, while the outer cylinder
remains fixed. The walls of the cylinders, in contact with the grains,
 are roughened by means of a glued layer of identical granular material,
thereby minimizing wall slip. The walls of the inner and outer cylinders are
roughened by acrylic beads with diameter $3.97$ and $1.59$mm, respectively.
Testing the experiment with a rough inner wall and a smooth outer wall resulted in packing
crystallization.  The grains are compacted by an external pressure
of a specific value (typically 386 Pa), introduced by a moving
piston at the top of the granular material, acting in the negative
z-direction.


\begin{figure}
\centerline{\resizebox{8cm}{!}{\includegraphics{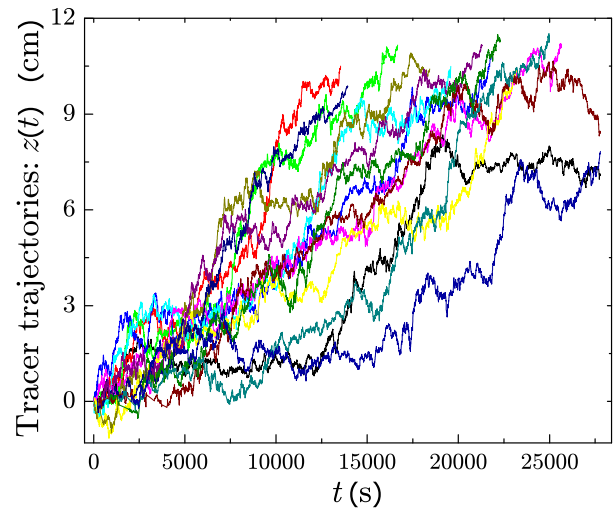}}}
\caption{(Color online) Trajectories of the $3.97$mm nylon tracers in Packing
1 showing the diffusion and response to the gravitational force
when sheared in the Couette cell.} \label{trajectory1}
\end{figure}

Observation techniques are used to monitor the granular packing
evolution as it explores the available jammed configurations. The
Couette cell is sheared at the quasi-static limit, with slow
frequencies $f=0.2\sim 4.2$ mHz defining the external shear-rate
$\dot \gamma_e = 2\pi f R_1/(R_2-R_1)=\omega
R_1/(R_2-R_1)=0.004\sim 0.084$ $\texttt{s}^{-1}$, where
$R_1=5.08$cm and $R_2=6.67$cm are the radius of the inner and
outer cylinders, respectively, and $\omega$ is the angular
velocity of the inner cylinder (Notice that $R_1$ and $R_2$ are measured after the walls are roughened by a glued layer of beads).  The experiment is designed to
measure the diffusivity and mobility of tracer particles
\cite{mk,swm,pch}, as opposed to tracking the motion of all
constituent grains. The distance between the inner and outer
cylinder is less than 10 grain diameters to prevent bulk shear
band formation \cite{nedderman,drake,mueth,veje,utter} that may
interfere with the experimental measurements by altering the
diffusivity.

A refractive index matching suspending solution is employed in
order to create a transparent sample.  The suspending solution is
also density matched to the grains in order to eliminate pressure
gradients derived from gravity in the vertical direction,
circumventing problems seen in previous experiments of
compactivity \cite{nowak} and other effects such as convection and
size segregation such as the Brazil nut effect inside the cell
\cite{behringer}. The solution used in this experiment is
approximately 74\% weight fraction of cyclohexyl bromide and 26\%
decalin \cite{weeks}. These steps avoids problems encountered in
previous tests of compactivity.



\subsection{Packing Preparation}\label{DRIM}

\begin{figure}
\centerline{\resizebox{5cm}{!}{\includegraphics{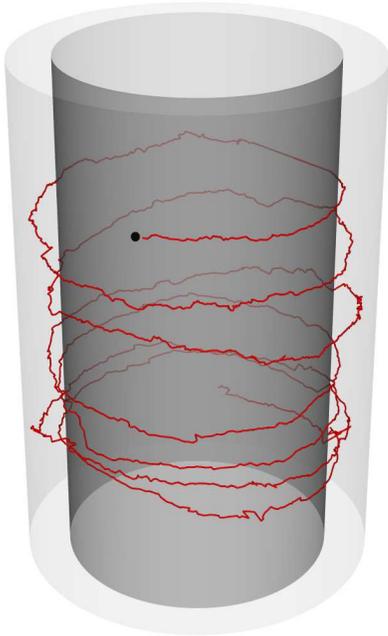}}}
\caption{(Color online) A typical trajectory of the $3.97$mm nylon tracer for
3 hours in 3D plot. The dark gray and light gray cylinder indicate
the outer surface of the sheared inner cylinder and the inner surface
of the static outer cylinder respectively.} \label{trajectory2}
\end{figure}

The granular system is a bidisperse, 1:1 by mass, mixture of
spherical, transparent Poly-methyl methacrylate (acrylic)
particles, with density $\rho = 1.19$ and index of refraction
$n\simeq1.49$. The bidisperse mixture is used in an effort to inhibit
crystallization of the system. The respective particle diameters
are either $3.17$mm and $3.97$mm (Packing 1) or $3.97$mm and
$4.76$mm (Packing 2). The approximate same size ratio of each
bidisperse packing leads to approximately the same value of volume
fraction for both, being $0.62$ before shearing and $0.58$ during
shearing.

\begin{figure}
\centering{\resizebox{8cm}{!}{\includegraphics{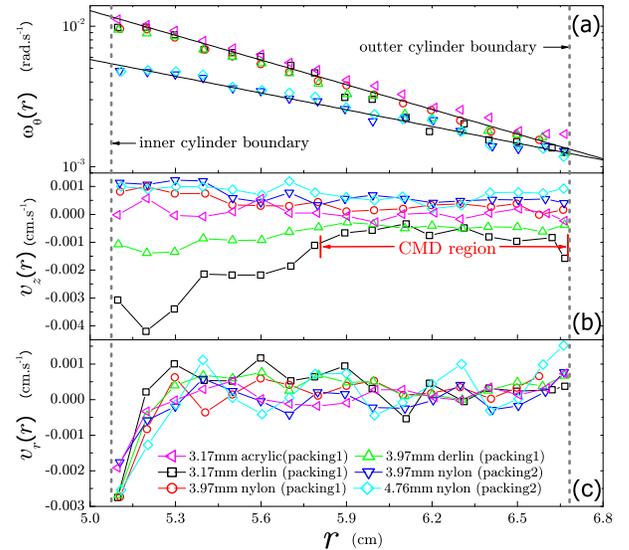}}}
\caption{(Color online) (a) Average angular velocity, $\omega_\theta(r)$, (b)
Average vertical velocity, $v_z(r)$, and (c) Average radial
velocity, $v_r(r)$, versus radial distance $r$ for various tracers
and different packings. Packing 1 and Packing 2 are run at
$\dot\gamma_e=0.048, 0.024\mathrm{s}^{-1}$, respectively. In (a), solid
 lines are
exponential fitting. In (b), the
positive velocity of nylon tracer is due to the smaller density
than acrylic's. The negative velocity of delrin tracer is due to
the higher density than acrylic's.} \label{V_z_t_r}
\end{figure}

A negative consequence of utilizing a suspending solution includes
possible modification of the friction coefficient between the
grains. While this cannot be completely avoided, it is important
to note that the liquid only partially fills the cell (see Fig.
\ref{setup1}), such that the pressure of the piston is transmitted
to the granular material exclusively, not to the fluid.
Additionally, hydrodynamic effects from partial cell filling are
avoided by the extremely slow rotational speeds applied to our
system.  The system remains very closely packed, such that
particles are not free to float in the fluid. Therefore, the
random motion of the particles is controlled by the `jamming'
forces exerted by the contacts between neighboring grains, not
fluid mechanics.


\subsection{Implementation of Fluctuation-Dissipation Theory}

\begin{figure}
\centering{\resizebox{8cm}{!}{\includegraphics{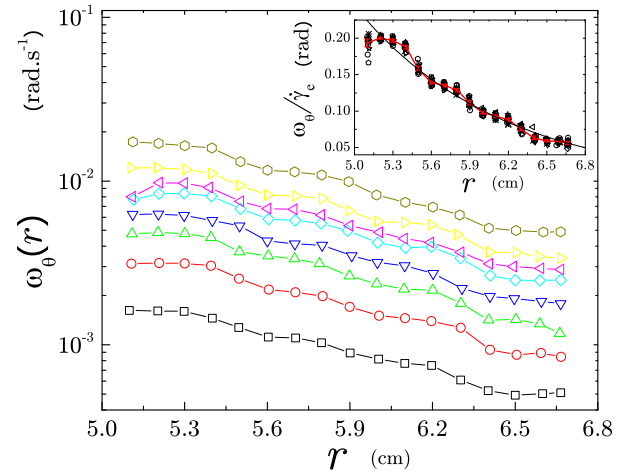}}}
\caption{(Color online) Average angular velocity of tracers, $\omega_\theta(r)$,
versus radial distance $r$ for various shear rate $\dot\gamma_e$
in Packing 2. Black square, red circle, green triangle, blue
triangle-down, cyan diamond, magenta triangle-left, yellow
triangle-right, dark yellow hexagon are corresponding to
$\dot\gamma_e=0.008,0.016,0.024,0.032,0.041,0.048,0.060,0.084\mathrm{s}^{-1}$,
respectively. The inset plots the collapsing of average angular
velocity scaled by shear rate, $\omega_\theta(r)/\dot\gamma_e$,
versus radial distance $r$ for various shear rate $\dot\gamma_e$.
The red solid curve is the average result of the collapsing. The
black solid curve is a exponential fitting.}
\label{Vtheta_r_shear}
\end{figure}

Cylindrical coordinates, $(r(t),\theta(t),z(t))$, of tracer
particles are obtained by analyzing images acquired by four
digital cameras surrounding the Couette cell.  For systems in
thermal equilibrium, a Fluctuation-Dissipation (FD) relation may
be utilized in an effort to calculate the bath temperature of the
system.  This method may be extended to non-equilibrium systems,
such as jammed granular systems presented in this study.  The FD
relation is defined as follows:

\begin{equation}
\langle [x(t+\Delta t) - x(t)]^2\rangle \sim 2 D \Delta t,
\end{equation}

\begin{equation}
\langle [x(t+\Delta t) - x(t)\rangle \sim M F \Delta t,
\end{equation}

\begin{equation}
\langle [x(t+\Delta t) - x(t)]^2\rangle = ~ 2 T_{\mbox{\scriptsize
eff}} ~ \frac{\langle x(t+\Delta t) - x(t)\rangle }{F}.
\label{fdt}
\end{equation}

The tracer particles must experience a constant force, $F$, in order to
calculate the mobility as defined above.  The most convenient
constant, external, force, is gravity in the z-direction.  If the
effective temperature is to be regarded as an intensive variable
of the non-equilibrium system, it requires independence from the
tracer particles properties, and we present data in favor of this
result. However, we acknowledge that temperature measures from
multiple observables would be necessary to analyze the underlying
thermodynamic meaning of $T_{\mbox{\scriptsize eff}}$.


\subsection{Properties of Tracer Particles}

Tracer particles added to the bulk must have properties unique
from the grains comprising the bulk.  However, tracer particles
too small, or too large, with respect to the acrylic grains
described previously, would result in erroneous measurements.
Dynamics of tracer particles that were too small would be
dominated by ``percolation effects'' \cite{drahun}, resulting in
larger than expected tracer particle displacements.  Those too
large would require shear rates above the quasi-static limit we
propose to study, or possibly have no dynamics at all due to size
limitations. With these notions in mind, two different types of
tracers, nylon ($\rho' = 1.12$) and delrin ($\rho' = 1.36$), are
employed, which result in different external forces,
$F=(\rho'-\rho)Vg$, where $\rho$ and
$\rho'$ are the densities of the acrylic particles and the
tracers, respectively, $V$ is the volume of the tracer particle
and $g$ the gravitational acceleration. Variations in tracer particle
diameter and density allow us to study dynamical changes due to a
change in constant external force, while we remain within a range
appropriate to achieve results expected to be governed by the
effective temperature.

\subsection{Particle Tracking Technique} \label{PTT}

Four digital cameras symmetrically surround the shear cell to
track the tracer particles with frame rate $\sim$ 5 frame/s, as
shown in Fig. \ref{setup3}. The outer cylinder is made of the same
material as the grains (acrylic, $n\simeq 1.49$). The refractive
index is matched by the fluid such that the system can be regarded
as an optical whole, i.e., the light scattered from tracers
refracts only once at the outer surface of the outer cylinder, as
shown in Fig. \ref{setup2}. The determination of the 3D tracer
position is achieved by a simple calculation considering both
system geometry and 2D projections captured by two adjacent
cameras.

\begin{figure}
\centering{\resizebox{8cm}{!}{\includegraphics{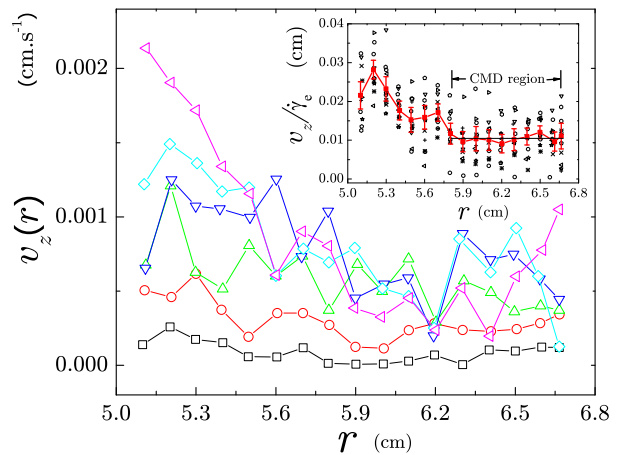}}}
\caption{(Color online) Average vertical velocity of tracers, $v_z(r)$, versus
radial distance $r$ for various shear rate $\dot\gamma_e$ in
Packing 2. Black square, red circle, green triangle, blue
triangle-down, cyan diamond, magenta triangle-left are
corresponding to
$\dot\gamma_e=0.008,0.016,0.032,0.048,0.060,0.084\mathrm{s}^{-1}$,
respectively. The inset plots the collapsing of average vertical
velocity scaled by shear rate, $v_z(r)/\dot\gamma_e$, versus
radial distance $r$ for various shear rate $\dot\gamma_e$. The red
solid curve is the average result of the collapsing.}
\label{Vz_r_shear}
\end{figure}

Camera calibration and determination of relative position is
important as a minimal asymmetry will result in a large
calculation error of the tracer particles coordinates,
$(r(t),\theta(t),z(t))$. As opposed to directly measuring relative
positions of the cameras by physical devices, we utilize computer
programming. In order to simplify the calculation, we assume the
camera to be a pinhole, meaning all light coming into the camera
coincides at a single focal point.

Before each experiment, we record the images of a piece of grid
paper attached to the surface of the outer cylinder, acting as the
2D projections of the outer cylinder for each camera. Next, we
adjust the positions of four cameras until each camera can give
the approximately same 2D projections of the outer cylinder. Then
we use the computer program to generate a virtual cylinder, along
with four virtual cameras, according to the geometry of the shear
cell, In other words, we build a virtual space of the entire
experimental setup and the respective geometrical relations
between its elements.

From the previous calibration procedure, we have the relative
positions of the four cameras to the shear cylinder with
sufficient accuracy. In order to further calibrate and know the
exact position of cameras, we adjust the relative position of
cameras in our virtual space until the virtual 2D projection of
the cylinder to the cameras coincides exactly with the actual
projection, being the grid paper attached to the outer cylinder.
When this procedure is accomplished, the virtual space exactly
coincides with the real experimental setup space.  Therefore the
virtual relative position of cameras are also the real positions.


Furthermore, in our virtual space, any point with 3D coordinates,
$(r,\theta,z)$, we can calculate its 2D coordinates in four
virtual 2D projections, $(x_1, y_1) \sim (x_4, y_4)$, by
considering the geometry relation to cameras. Oppositely, for any
tracer particle, if we know its 2D coordinates in four 2D
projections, $(x_1, y_1) \sim (x_4, y_4)$, we can exactly locate
its 3D positions, $(r,\theta,z)$, since the virtual space is equal
to the actual one. The resulting vertical trajectories of the
tracers $z(t)$ are depicted in Fig. \ref{trajectory1} showing
that the nylon tracers not only diffuse, but also move with a
constant average velocity to the top of the cell. Fig.
\ref{trajectory2} shows a typical trajectory of tracer particle
in 3D plotting.

\section{Results}
\subsection{Average Velocity Profiles}

We first study the velocity profiles for a fixed shear rate,
$\dot{\gamma_e}$, followed by a study on the shear rate dependence
in the next section. The average velocity profiles in the angular
direction, $\omega_\theta(r)$, in the vertical direction,
$v_z(r)$, and in the radial direction, $v_r(r)$, are obtained by
averaging the velocities of all tracer particles over all times at
each radius $r$, as shown in Fig. \ref{V_z_t_r}.

As observed in previous work \cite{mueth2}, we find that
$\omega_\theta(r)$ can be expressed in the exponential form
demonstrated in Fig. \ref{V_z_t_r}a:

\begin{equation}
\begin{split}
\omega_\theta(r)&= \lambda_1\frac{\dot\gamma_e(R_2-R_1)}{R_1}
\exp(-\lambda_2 \frac{r-R_1}{R_2-R_1})\\
&= \lambda_1\omega\exp(-\lambda_2 \frac{r-R_1}{R_2-R_1}),
\end{split}
\label{vtheta_exp}
\end{equation}
where $\lambda_1$ and $\lambda_2$ are constants independent of
shear rate, tracer size and tracer type, depending only on the
type of packings and geometry of the shear cell. We find that
$\lambda_1=0.77$ and $0.73$, and $\lambda_2=2.15$ and $1.43$ for
Packing 1 and Packing 2, respectively. When $r=R_1$,
$\omega_\theta(R_1)=\lambda_1\omega$ being the angular velocity of
the first layer of grains closest to the sheared inner cylinder
with angular velocity $\omega$. Therefore $\lambda_1$
($0<\lambda_1<1$) can be taken as the efficiency of shearing,
describing the amount of slip between the inner rotating cylinder
and the first layer of grains it contacts. Packing 1 has a higher
value of $\lambda_1$ than Packing 2, as the smaller grains follow
the rotating inner cylinder more easily. On the other hand, when
$r=R_2$, we find
$\omega_\theta(R_2)=\lambda_1\omega\exp(-\lambda_2)$, the velocity
of the last layer of grains closest to the static outer cylinder.
This velocity is non-zero, so that the shear band is located right
at the outer cylinder, avoiding the formation of shear bands in
the bulk.

\begin{figure}
\centering{\resizebox{8cm}{!}{\includegraphics{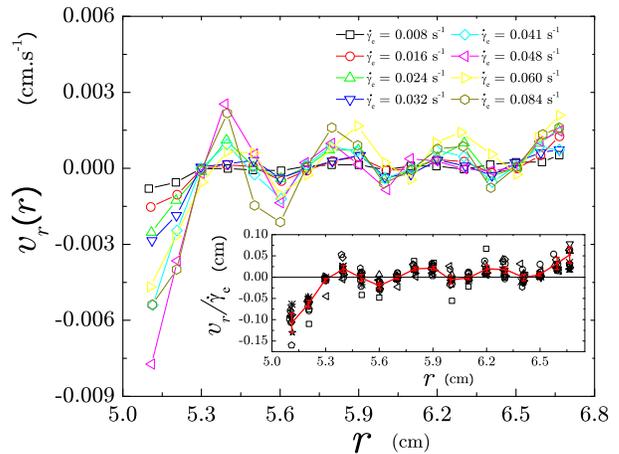}}}
\caption{(Color online) Average radial velocity of tracers, $v_r(r)$, versus
radial distance $r$ for various shear rate $\dot\gamma_e$ in
Packing 2. Black square, red circle, green triangle, blue
triangle-down, cyan diamond, magenta triangle-left, yellow
triangle-right, dark yellow hexagon are corresponding to
$\dot\gamma_e=0.008,0.016,0.024,0.032,0.041,0.048,0.060,0.084\mathrm{s}^{-1}$,
respectively. The inset plots the collapsing of average radial
velocity scaled by shear rate, $v_r(r)/\dot\gamma_e$, versus
radial distance $r$ for various shear rate $\dot\gamma_e$. The red
solid curve is the average result of the collapsing.}
\label{Vr_r_shear}
\end{figure}

In order to avoid the tracer particles sticking to the outer
cylinder surface and forcing its velocity to zero, we glue smaller
size particles to roughen the outer cylinder.  This roughens the
surface of the outer cylinder and avoids crystallization. Further,
this allows slipping of the bulk particles at the outer cylinder,
forcing the shear band to be located exactly at the outer
cylinder, not in the bulk.  The glued particles are $1.59$mm,
smaller than the sheared granular material. The mean angular and
vertical velocity, $\omega_\theta$ and $v_z$, of tracer particles
do not decay to zero even if the tracers come close to outer
cylinder surface. (See Fig. \ref{Vtheta_r_shear} and Fig.
\ref{Vz_r_shear} at $r=R_2$).

The exponential decay of $\omega_\theta(r)$ results in a local
shear rate, $\frac{d\omega_\theta(r)}{dr}$, dependent on radial
distance.  In the region near the outer cylinder,
$\omega_\theta(r)$ decays slowly with increasing $r$ which leads
to weak dependence of $\frac{d\omega_\theta(r)}{dr}$ on the radial
distance $r$. If we do Taylor expansion at $r=R_2$, the average angular
 velocity of the tracers,
$\omega_\theta(r)$, can be approximated to a linear function of $r$,
 i.e.,
$\omega_\theta(r)\approx\lambda_1\omega\exp(-\lambda_2)-r\frac{d\omega_\theta(r)}{dr}$
with constant local shear rate $\frac{d\omega_\theta(r)}{dr} =
0.021$ s$^{-1}$cm$^{-1}$. The diffusivity and mobility of the tracer particles strongly depend on the local rearrangement of the grains. A constant shear rate results in homogenous local rearrangement of the packings ensuring that the diffusivity and mobility of tracers, dependent on local shear rate, remain approximately independent of $r$. As shown in
Fig. \ref{V_z_t_r}b, we find a plateau in the vertical velocity
profile which can be further seen in Fig. \ref{Vz_r_shear}.
Similar behavior is observed in the vertical diffusivity profile,
$D_z(r)$ as shown in Fig. \ref{D_r_shear}, which we will discuss in
 detail in
Section \ref{Teff}. We denote this the ``constant
mobility and diffusivity region'', i.e., CMD region, $5.80$cm $< r
< 6.67$cm. Contrary to prior work
\cite{howell,veje,mueth,mueth2,utter,nedderman,drake} on sheared
granular matter in the Couette cell, our experiment focuses only a
narrow gap, $15.9$mm, of Couette cell.  The CMD region allows us
to well define the diffusivity and mobility of the tracer
particles, such that we can calculate the average vertical
velocity, $v_z$, and the average vertical diffusivity, $D_z$, by
averaging the velocities of all tracers over all times in the CMD
region, significantly improving the statistics. In this study, the
statistical average and the measurements of tracer fluctuations
will be confined only to the CMD region.


\begin{figure}
\centering{\resizebox{8cm}{!}{\includegraphics{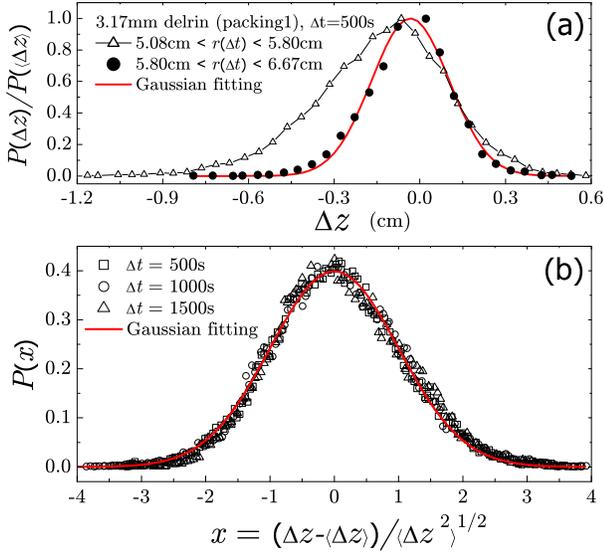}}}
\caption{(Color online) (a) PDF of the vertical displacements, $P(\Delta z)$, of
the $3.17$mm delrin tracers in Packing 1 for a given time interval
$\Delta t=50$s, and with $\dot\gamma_e=0.048\mathrm{s}^{-1}$. Tracer
trajectories are split into sub-trajectories confined in two
regions, (i): $5.08$cm $< r < 5.80$cm, which is close to inner
rotating cylinder, and (ii): $5.80$cm $< r < 6.67$cm, which is far
away from inner rotation cylinder. We compared the calculated
$P(\Delta z)$ by using the sub-trajectories from the regions of
(i) and (ii) respectively, which are plotted as black triangle and
black circle. See more details in the main text. (b) PDF of the
vertical displacements, $P(\Delta z)$, of the $3.97$mm nylon
tracers in Packing 1 with $\dot\gamma_e=0.048\mathrm{s}^{-1}$, shifted
 by
the average displacement $\langle\Delta z\rangle$ and scaled by
the root-mean-square deviation $\langle\Delta
z(t)^2\rangle^{1/2}$. The red solid curve is a Gaussian
distribution, $P(x)=0.4\text{e}^{-x^2/2}$.} \label{Pz}
\end{figure}

\begin{figure}
\centerline{\resizebox{8cm}{!}{\includegraphics{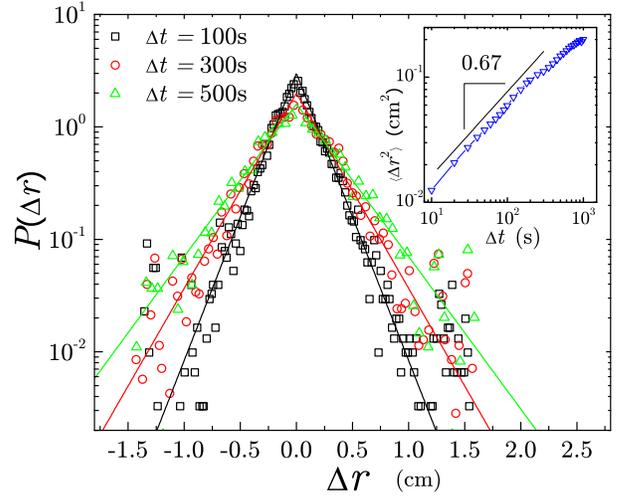}} }
\caption{(Color online) PDF of the radial displacements, $P(\Delta r)$, of the
$3.97$mm nylon tracers in Packing 1 with
$\dot\gamma_e=0.048\mathrm{s}^{-1}$ for different time intervals. A
symmetric distribution around zero displacement indicates that
there is no net flow in the radial direction. The solid lines are
exponential fitting, $P(\Delta r)\sim\text{exp}(-\frac{|\Delta r|}{r_o})$,
where $r_o=0.17, 0.25, 0.32$ for $\Delta t=100,300,500$s, respectively. The
inset shows the rms fluctuations, which gives the value of
$\alpha=0.67$.} \label{Pr}
\end{figure}

We find $v_r(r)$ to be flat for different types of tracer
particles and for different packings except when the tracers are
close to the inner and outer cylinder, i.e., $r=R_1$ and $r=R_2$,
as shown in Fig. \ref{V_z_t_r}c. $v_r(R_1)$ is negative and
$v_r(R_2)$ is positive, indicating the inner and outer cylinder
walls can slightly attract the tracers. It should be noted that
the statistics presented in this study does not incorporate data
from the regions close to the inner and outer cylinder to avoid
these boundary effects.

\begin{figure}
\centerline{\resizebox{8cm}{!}{\includegraphics{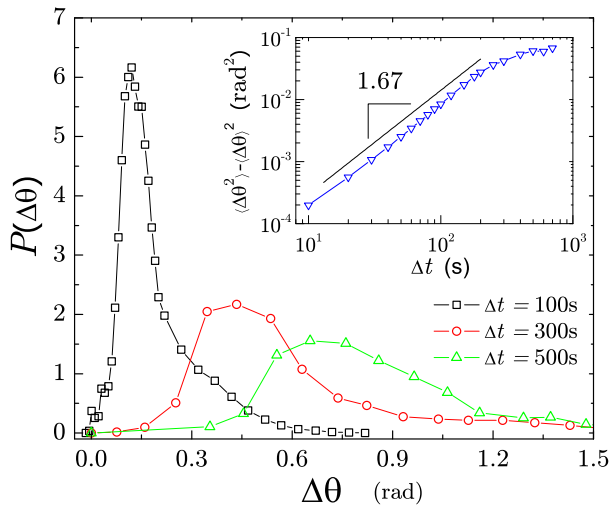}}}
\caption{(Color online) PDF of the angular displacements, $P(\Delta \theta)$, of
the $3.97$mm nylon tracers in Packing 1 with
$\dot\gamma_e=0.048\mathrm{s}^{-1}$ for different time intervals. Due
 to
Taylor dispersion effects the distribution shows an asymmetric
shape. The rms fluctuations shown in the inset reveal a faster
than diffusion process.} \label{Ptheta}
\end{figure}

\subsection{Shear Rate Dependent Average Velocity Profiles}

Next, we study the dependence of the particle velocity on the
external shear rate.  According to Eq. (\ref{vtheta_exp}), the
velocity profile in the angular direction, $\omega_\theta(r)$, is
proportional to the external shear rate $\dot\gamma_e$. We can
collapse $\omega_\theta(r)$ by scaling the shear rate. The results
are shown in the inset of Fig. \ref{Vtheta_r_shear} for Packing 2.
The collapsing of $\omega_\theta(r)/\dot\gamma_e$ shows a periodic
shape superimposed to exponential decay with a very small
amplitude, also found in the velocity profile of $v_r(r)$ (see
Fig. \ref{Vr_r_shear}). The periodic length is roughly equal to
the grain particle size and reflects the different layers of
grains in the radial direction. This periodicity is weaker in
Packing 1 than Packing 2, since the particle size of Packing 1 is
smaller than that of Packing 2.

The collapsing method can be further applied to $v_z(r)$, as seen
in Fig. \ref{Vz_r_shear}. After scaling by the shear rate,
$v_z(r)/\dot\gamma_e$ also shows a flat plateau indicating the CMD
region.

\subsection{Probability Distribution of Displacements}

Fig. \ref{Pz}a  shows the results of the probability distribution
of the displacements $\Delta z$ in the vertical direction for a
given time interval $\Delta t$. The data corresponds to the
$3.17$mm delrin tracers in Packing 1. Usually, 20 tracers are used
for calculations.  Tracer trajectories are split into
sub-trajectories confined in two regions, (i): $5.08$cm $< r <
5.80$cm, close to the inner rotating cylinder, and (ii): $5.80$cm
$< r < 6.67$cm, i.e., CMD region, close to the outer cylinder. We
compare the calculated $P(\Delta z)$ by using the sub-trajectories
from the regions of (i) and (ii) respectively, which are plotted
as black triangle and black circle in the Fig. \ref{Pz}a. The data
in the inner region (i) clearly display an asymmetric tail for
$\Delta z <0$. This extra spreading is similar to the phenomena of
the Taylor dispersion \cite{taylor}.

\begin{figure}
\centerline{\resizebox{8cm}{!}{\includegraphics{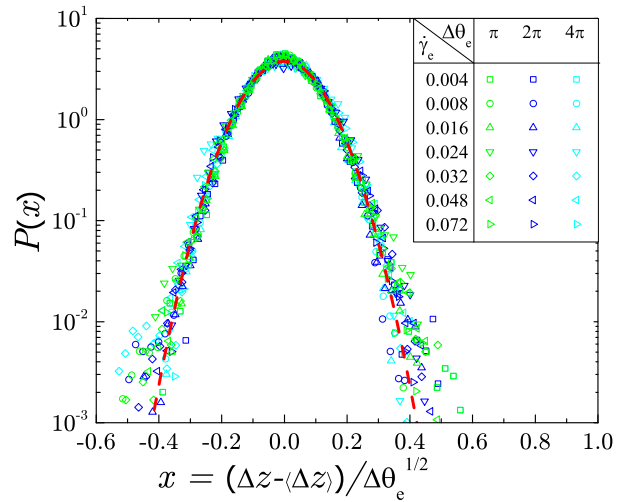}}
} \caption{(Color online) PDF of the vertical displacements, $P(\Delta z)$, of
the $4.76$mm nylon tracers in Packing 2 for various effective
angular displacement $\Delta{\theta_e}$ and effective shear rate
$\gamma_e$. The PDFs are scaled by $\Delta{\theta_e}^{1/2}$ and
shifted by the mean displacement $\langle\Delta z\rangle$. The red
dashed line is the Gaussian fitting, $P(x)\sim \text{exp}[-(\frac{x}{0.147})^2]$.
The collapsing of PDFs indicates that the RMS fluctuations of the vertical displacements
follow the relation, $\langle\Delta z^2\rangle\sim\Delta{\theta_e}$.} \label{Pz_collapsing}
\end{figure}

Taylor dispersion appears when diffusion couples with the gradient
of flow giving rise to a larger dispersion along the flowing
direction (see for instance \cite{utter} for a study of Taylor
dispersion in granular materials). In the present experiment, the
shear rate of granular flow in the angular direction exhibits
exponential decay, as shown in Fig. \ref{V_z_t_r}b. The larger
shear rate in the inner region (i) results in larger packing
rearrangement, which gives rise to a larger dispersion in the
vertical direction. In this case it is not possible to extract the
bare diffusion constant. On the contrary, for the region (ii),
i.e., CMD region, as we mentioned, the gradient of the flow, i.e.,
the shear rate is approximately constant, giving rise to a
Gaussian diffusion, as shown in Fig. \ref{Pz}a. By measuring the
width and the mean value of this Gaussian distribution of

\begin{equation}
\begin{split}
P(\Delta z)&\sim\exp[{-\frac{(\Delta z-\langle\Delta
z\rangle)^2}{2\langle \Delta z^2\rangle}}] \\
&\sim\exp[{-\frac{(\Delta z-M_z F \Delta t)^2}{4D_z \Delta t}}],
\end{split}
\end{equation}
we can define the
diffusivity and mobility, $D_z$ and $M_z$, which lead to the
effective temperature of the granular packing discussed in the
following section. In the Fig. \ref{Pz}b, we define a new scaled
variable $x=\frac{\Delta z-\langle\Delta z\rangle}{{\langle \Delta
z^2\rangle}^{1/2}}$ and plot $P(x)$ for different $\Delta t$, all
the curves are found to collapse into a single curve

\begin{equation}
P(x)\sim\text{e}^{-x^2/2}.
\end{equation}
In this experiment, we will focus our
measurements in the region away from the inner boundary (region
(ii), i.e., CMD region), where the mobility is a constant (as
shown a plateau in the inset of Fig. \ref{Vz_r_shear}) and Taylor
dispersion effects are absent.

\begin{figure}
\centerline{\resizebox{8cm}{!}{\includegraphics{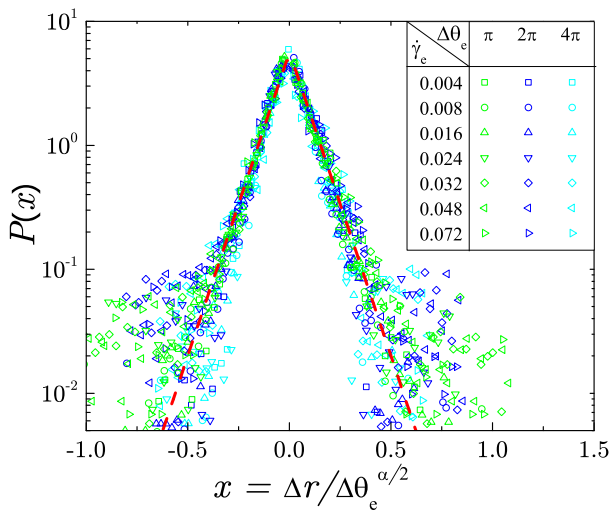}}
} \caption{(Color online) PDF of the radial displacements, $P(\Delta r)$, of the
$4.76$mm nylon tracers in Packing 2 for various effective angular
displacement $\Delta{\theta_e}$ and effective shear rate
$\gamma_e$. The PDFs are scaled by $\Delta{\theta_e}^{\alpha/2}$,
where $\alpha=0.67$. The red dashed line is exponential fitting,
$P(x)\sim \text{exp}(-\frac{|x|}{0.089})$.
The collapsing of PDFs indicates that the RMS fluctuations of the
radial displacements follow the relation, $\langle\Delta
r^2\rangle\sim\Delta{\theta_e}^{\alpha}$.} \label{Pr_collapsing}
\end{figure}

We find exponential fluctuations for the probability distributions of
 the
tracer particles in the radial direction as shown in Fig.
\ref{Pr},

\begin{equation}
P(\Delta r)\sim\text{e}^{-\frac{|\Delta r|}{r_o}},
\end{equation}
where $r_o$ is a function of $\Delta t$. The symmetric shape for $P(\Delta r)$
 indicates the
absence of a shear induced segregation, as observed with multiple
sizes of grains, as there is no net flow of the tracer particles
towards either cylindrical wall within the time-scales of the
experiment.  We also observe no average motion of the tracer
particles towards the center of the Couette cell except within a
small range of radial distance, around $0.12$cm, close to both
walls where particles experience a slight attraction to
boundaries.  These features are shown in Fig. \ref{V_z_t_r}c and
Fig. \ref{Vr_r_shear}.


The analysis of the radial displacement fluctuation reveals a
power law, sub-diffusive, process:
\begin{equation}
\langle\Delta r^2\rangle\sim{\Delta t}^\alpha
\end{equation}
as shown in Figure 8, where $\alpha=0.67$ for both the delrin and
nylon tracers.


The data taken for the angular displacement is in the direction of
the flow, and affected by Taylor dispersion as shown in the
non-Gaussian tail of the displacement distribution
$\Delta\theta(t)$ in Fig. \ref{Ptheta}. This leads to a power law,
super-diffusive process, illustrated by
\begin{equation}
\langle\Delta\theta^2\rangle\sim{\Delta t}^\beta
\end{equation}
as seen in the analysis of the fluctuations of $\Delta \theta$
shown in Fig. \ref{Ptheta}, where $\beta=1.67$ and $1.30$ for
Packing 1 and Packing 2 respectively.

We further study how shear rate affects the displacement
probability distribution. We find that for small shear rate, the
probability distributions of displacements in the three
cylindrical coordinates are independent of the shear rate,
depending only on the sheared displacement, i.e., the external
rotating displacement, defined as

\begin{equation}
\begin{split}
\Delta\theta_e &= \dot\gamma_e\Delta t \\
&= \omega\Delta tR_1/(R_2-R_1) \\
&= \Delta\theta_iR_1/(R_2-R_1)
\end{split}
\end{equation}
where $\Delta\theta_i$ is the rotating displacement of the inner
cylinder. This result is expected. Since we shear the Couette cell
very slowly, the diffusion of the tracers depends only the number
of granular packing configurations sampled by the Couette cell,
which depends only on the sheared displacement.

As emphasized in the previous text, the statistical average and
the measurements of the tracer fluctuations is confined to the CMD
region, such as the $D_z$ shown in Fig. \ref{D_M_T}a. We calculate
the $D_z$ by measuring the width and the mean value of the
Gaussian distribution of $P(\Delta z)$, and obtain the $P(\Delta
z)$ by averaging the displacement fluctuations of all tracers over
all time in the CMD region. Next, we apply a different method to
reveal how $D_z(r)$ depends on the radial distance $r$, as shown
in the Fig. \ref{D_r_shear}. We first obtain $P(\Delta z, r)$ for
a certain radial distance $r$, then we calculate $D_z(r)$ by
measuring the width and the mean value of the Gaussian
distribution of $P(\Delta z, r)$. In Fig. \ref{D_r_shear}, we see
that the tracer particles have higher diffusivity close to the
inner cylinder than the outer. Since $D_z\sim \dot\gamma_e$, we
can collapse all the $D_z(r)$ for various shear rates, as shown in
the inset of Fig. \ref{D_r_shear}. The collapse of
$D_z(r)/\dot\gamma_e$ shows a plateau close to the outer cylinder,
consistent with our previous discussion of the CMD region.

\begin{figure}
\centerline{\resizebox{8cm}{!}{\includegraphics{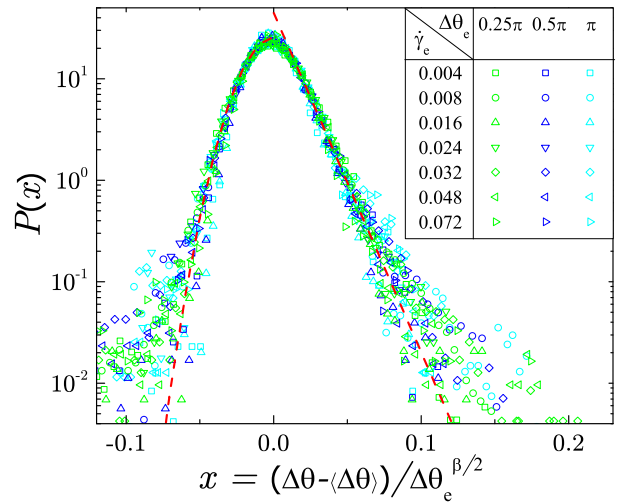}}
} \caption{(Color online) PDF of the angular displacements, $P(\Delta \theta)$,
of the $4.76$mm nylon tracers in Packing 2 for various effective
angular displacement $\Delta{\theta_e}$ and effective shear rate
$\gamma_e$. The PDFs are shifted by the mean displacement
$\langle\Delta \theta\rangle$ and scaled by
$\Delta{\theta_e}^{\beta/2}$, where $\beta=1.30$. The red dashed
line are Gaussian and exponential fittings for $x<0$ and $x>0$,
respectively. The collapsing of PDFs indicates that the rms
fluctuations of the angular displacements follow the relation,
$\langle\Delta \theta^2\rangle\sim\Delta{\theta_e}^{\beta}$.}
\label{Ptheta_collapsing}
\end{figure}


\begin{subequations}\label{fluctuations}
\begin{align}
\langle[z(t+\Delta t)-z(t)]^2\rangle &\sim\Delta\theta_e \\
\langle[r(t+\Delta t)-r(t)]^2\rangle &\sim{\Delta\theta_e}^\alpha \\
\langle[\theta(t+\Delta t)-\theta(t)]^2\rangle
&\sim{\Delta\theta_e}^\beta
\end{align}
\end{subequations}
Eq. (\ref{fluctuations}) implies that one can collapse the
probability distribution of the displacements, $P(\Delta z)$,
$P(\Delta r)$ and $P(\Delta\theta)$ for different shear rates and
time intervals by scaling $\Delta z$, $\Delta r$ and
$\Delta\theta$ respectively to $\Delta z/{\Delta\theta_e}^{1/2}$,
$\Delta r/{\Delta\theta_e}^{\alpha/2}$ and
$\Delta\theta/{\Delta\theta_e}^{\beta/2}$. The results are
presented in Fig. \ref{Pz_collapsing}, \ref{Pr_collapsing} and
\ref{Ptheta_collapsing}.

\begin{figure}
\centering{\resizebox{8cm}{!}{\includegraphics{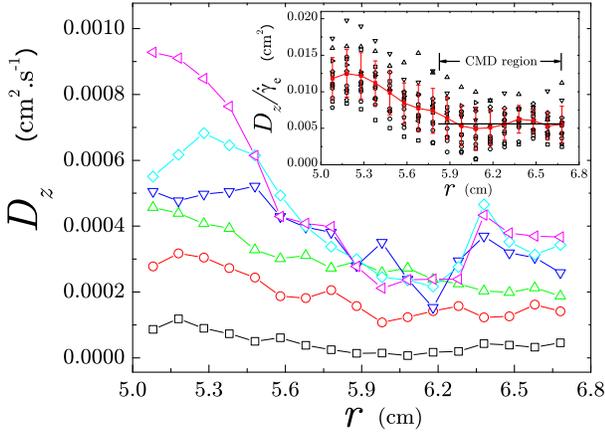}}}
\caption{(Color online) Diffusivity $D_z$ versus radial distance $r$ for various
shear rate $\dot\gamma_e$ in Packing 2. Black square, red circle,
green triangle, blue triangle-down, cyan diamond, magenta
triangle-left are corresponding to
$\dot\gamma_e=0.008,0.016,0.032,0.048,0.060,0.084\mathrm{s}^{-1}$,
respectively. The inset plots the collapsing of diffusivity scaled
by shear rate, $D_z/\dot\gamma_e$, versus radial distance $r$ for
various shear rate $\dot\gamma_e$. The red solid curve is the
average result of the collapsing.} \label{D_r_shear}
\end{figure}

\subsection{Effective Temperature} \label{Teff}

\begin{table*}
\begin{tabular}{|c|c|c|c|c|c|}
\hline
\multicolumn{6}{|c|}{Packing 1, 1:1 mass mixture of $3.17$mm \& $3.97$mm
 acrylic beads, $\dot\gamma_e=0.048\mathrm{s}^{-1}$}\\
\hline
tracer&$d$ \texttt{\scriptsize{(mm)}}&$\rho$
 \scriptsize{(g$\cdot$cm$^{-3}$)}&$D_z$ \scriptsize{($10^{-8}$ m$^2$$\cdot$s$^{-1}$)}&$M_z$
 \scriptsize{($10^{-2}$ s$\cdot$kg$^{-1}$)}&$T_\mathrm{eff}$
 \scriptsize{($10^{-7}$ J)}\\
\hline
acrylic&$3.17$&$1.19$&$2.5\pm 0.3$&&\\
\hline
delrin&$3.17$&$1.36$&$2.4\pm 0.3$&$24\pm 3$&$1.0\pm0.2$\\
\hline
delrin&$3.97$&$1.36$&$1.2\pm 0.1$&$9.3\pm 0.9$&$1.3\pm0.2$\\
\hline
nylon&$3.97$&$1.12$&$1.1\pm 0.1$&$9.5\pm 0.9$&$1.2\pm0.2$\\
\hline
ceramic&$3.97$&$3.28$&&$2.2\pm 0.2$&\\
\hline
brass&$3.97$&$8.4$&&$1.7\pm 0.1$&\\
\hline
\multicolumn{6}{|c|}{Packing 2, 1:1 mass mixture of $3.97$mm \& $4.76$mm
 acrylic beads, $\dot\gamma_e=0.024\mathrm{s}^{-1}$}\\
\hline
tracer&$d$ \texttt{\scriptsize{(mm)}}&$\rho$
 \scriptsize{(g$\cdot$cm$^{-3}$)}&$D_z$ \scriptsize{($10^{-8}$ m$^2$$\cdot$s$^{-1}$)}&$M_z$
 \scriptsize{($10^{-2}$ s$\cdot$kg$^{-1}$)}&$T_\mathrm{eff}$
 \scriptsize{($10^{-7}$ J)}\\
\hline
nylon&$3.97$&$1.12$&$1.8\pm 0.1$&$19.0\pm 0.9$&$0.95\pm0.07$\\
\hline
nylon&$4.76$&$1.12$&$1.6\pm 0.1$&$15.7\pm 0.4$&$1.0\pm0.1$\\
\hline
\end{tabular}
\caption{Diffusivity and mobility for the different types of
tracer and packings.}\label{table}
\end{table*}

We present results for the diffusivity in the $z$ direction, the
only direction where the effective temperature can be calculated
due to the vertically acting external force.  The Gaussian
distribution in $P(\Delta z)$ allows us to apply the FD relation
to the particle displacements, as the diffusivity is proportional
to the variance of a Gaussian distribution in displacements.
Exponential fluctuations do not possess this same property, but it
is important to note that the radial direction has no constant
applied external force. It remains a possibility that a
well-defined effective temperature for displacements in the radial
direction could exist.  To test whether the effective temperature
is isotropic, as done in \cite{danna}, may be of great interest in
future studies.


A common method of performing a time average to measure transport
coefficients is employed (see Chapter 5.3 in \cite{rapaport}) by
dividing the trajectory of a single tracer particle into a series
of trajectories, having evenly spaced start times, separated by
time interval $\Delta t$. The diffusion constant is obtained by averaging
over the aggregate of tracers and over the initial time intervals,
allowing for the use of merely 20 tracer particles in this
particular system. Correlations between measurements are ensured
to have decayed almost to zero, rendering time-translational
invariance valid in this system, without any measurable``aging'',
since under shearing, system reaches the ``stationary state''
\cite{coniglio}. Furthermore, doubling the number of tracer
particles leaves $D_z$ unchanged, indicating independence of the
diffusion constant from the number of tracers that explore the
jammed configurations of this non-equilibrium system.

\begin{figure}
\centering{\resizebox{8cm}{!}{\includegraphics{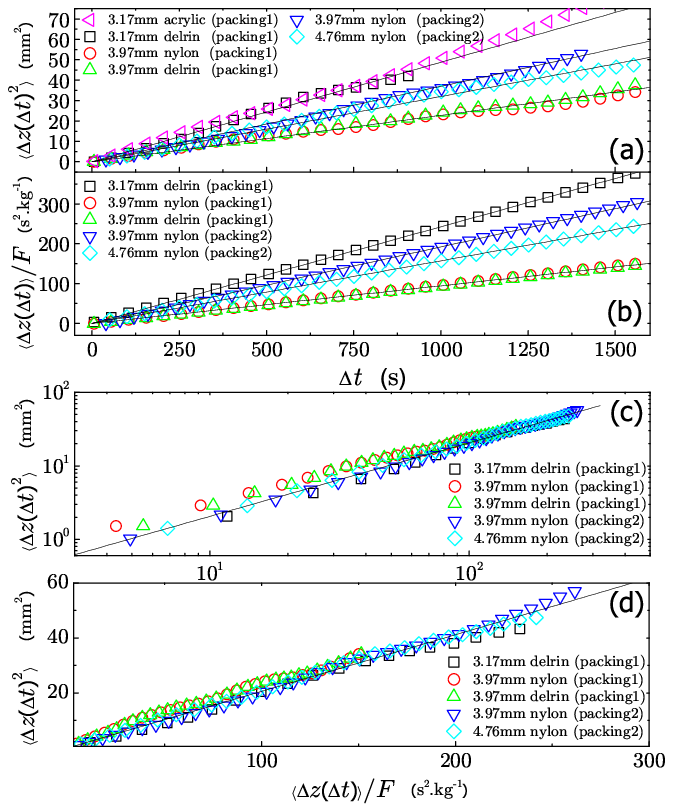}}}
\caption{(Color online) (a) Autocorrelation function of tracers. (b) Response
function of tracers. (c) Log-log plot of effective temperatures for various
tracers and different packings as obtained from a parametric plot
of their autocorrelation function versus response function.
(d) Same as (c) but in a linear-linear plot.
The slopes for different
tracer diffusivity vs. mobility curves return the same average
value of $T_{\mbox{\scriptsize eff}} \approx (1.1 \pm 0.2) \times
10^{-7}$J as given by Eq. (\ref{fdt}).} \label{D_M_T}

\end{figure}

Analysis of the vertical particle displacements in the CMD region
reveals a Gaussian distribution, broadening over time, as seen in
Fig. \ref{Pz} and Fig. \ref{Pz_collapsing}. For sufficiently long
times period, the mean square fluctuations grow linearly (see Fig.
\ref{D_M_T}a):
\begin{equation}
\langle[z(t+\Delta t) - z(t)]^2\rangle \sim 2D_z\Delta t,
\end{equation}
where $D_z$ is the self-diffusion constant in the vertical
direction. For the both nylon and delrin $3.97$mm tracers
in Packing 1 we obtain $D_{z\
\mbox{\scriptsize 3.97mm}} \approx ( 1.15 \pm 0.1 )\times 10^{-8}$
m$^2$/s.

Figure \ref{D_M_T}b shows mean value tracer particle positions,
extracted from the peak of the Gaussian distribution, as a
function of time. The mobility in the vertical direction,
$M_z$, is defined as
\begin{equation}
\langle z(t+\Delta t) - z(t)\rangle \sim M_zF\Delta t.
\end{equation}
The applied force on the tracers, $F = (\rho - \rho') V g$, is the
gravitational force due to density mismatch where $\rho$ and
$\rho'$ are the densities of the acrylic particles and the
tracers, respectively, $V$ is the volume of the tracer particle
and $g$ the gravitational acceleration. The value of the mobility
for the both nylon and delrin $3.97$mm tracers
in Packing 1 is $M_{z\ \mbox{\scriptsize 3.97mm}}\approx
( 9.4 \pm 0.9)\times 10^{-2} $ s/kg.

Fig. \ref{D_M_T}a further reveals a downward curvature of the
mean-square fluctuations, for sufficiently long times period.
Additionally, an apparent cut-off time for the tracer particles
fluctuation measurements is shown.  These effects are due to the
finite size effect imposed upon the tracers by the finite
trajectories and should be inversely proportional to the tracer
particles velocities.  Tracer particles with larger mobility will
have larger mean velocities and take a shorter time to complete
its trajectory in the cell.  The cut-off discussed in reference to
Fig. \ref{D_M_T}a is prominently displayed in the $3.17$mm delrin
tracers of Packing 1, having the largest mobility, hence increased
mean velocities, as shown in Fig. \ref{D_M_T}b.  The larger
mobility results in the shortest cut-off time for the diffusivity.
Conversely, $3.97$mm delrin tracers of Packing 1 have a smaller
mobility, hence a longer cut-off time for the diffusivity.  It is
important to note that for all tracer particles studied here, the
cut-off is observed for distances larger than a few particles
diameters, ensuring that the study examines the structural motion
of the grains and not internal motion inside of ``cages''.



According to a Fluctuation-Dissipation relation, we calculate
$T_{\mbox{\scriptsize eff}}$:

\begin{equation}
T_{\mbox{\scriptsize eff}} = \frac{F\langle [z(t+\Delta t) -
z(t)]^2\rangle}{2\langle z(t+\Delta t) - z(t)\rangle }.
\label{tcalc}
\end{equation}

Fig. \ref{D_M_T}c shows a parametric plot of fluctuations and
responses, with $\Delta t$, as the parameter, as extracted from Fig.
\ref{D_M_T}a and Fig. \ref{D_M_T}b.  A linear relationship exists
between diffusivity and mobility, with a slope of
$T_{\mbox{\scriptsize eff}}$. We obtain for the both nylon and delrin
 $3.97$mm tracers in Packing 1, $T_{\mbox{\scriptsize eff}} \approx
 (1.25 \pm 0.2) \times 10^{-7} \text{J}$.

If the effective temperature is to be regarded as an intensive
thermodynamic quantity, changing the tracer particle size should
give rise to a different diffusion and mobility yet result in the
same measurement of effective temperature. The above calculation
is repeated for delrin tracers of $3.17$mm in Packing 1. We find that while
the mobility and diffusivity change dramatically with respect to
tracers of $3.97$mm, ($D_{z\ \mbox{\scriptsize 3.17mm}} = ( 2.4
\pm 0.3)\times 10^{-8}$ m$^2$/s and $M_{z\ \mbox{\scriptsize
3.17mm}} = (2.4 \pm 0.3) \times 10^{-1}$ s/kg) as shown in
Table \ref{table}, due to the change
in tracer size, their ratio remains unchanged. In all cases $D_z$
and $M_z$ are inversely proportional to the size of the tracers,
but the effective temperature remains approximately the same, as
seen in Fig. \ref{D_M_T}c,  with an average value over all tracers
of

\begin{equation}
T_{\mbox{\scriptsize eff}}
\approx (1.1 \pm 0.2) \times 10^{-7} \text{J}.
\end{equation}

Though this effective temperature is high with respect to the bath
temperature, we note that a plausible scale for the system energy
\cite{review}, is $(\rho - \rho') g d$, the gravitational
potential energy to move a nylon tracer particle one particle
diameter, $d$. A corresponding temperature would arise from the
conversion of this energy into a temperature via the Boltzmann
constant, $k_B$, is $T_{\mbox{\scriptsize eff}} = 2.7 \times
10^{13} k_B T$ at room temperature (T = 300K). This specific value
serves as a coarse-grained estimate, since the tracer size and
density clearly shift its value, and we focus on the order of
magnitude. This large value is expected \cite{review}, and agrees
with computer simulation estimates for an athermal granular system
\cite{mk}. Therefore, our calculated value for
$T_{\mbox{\scriptsize eff}}$ in a sheared granular system appears
reasonable within the boundaries of the present theory.



\subsection{Linear Response Regime} \label{LRes}

\begin{figure}
\centering{\resizebox{8cm}{!}{\includegraphics{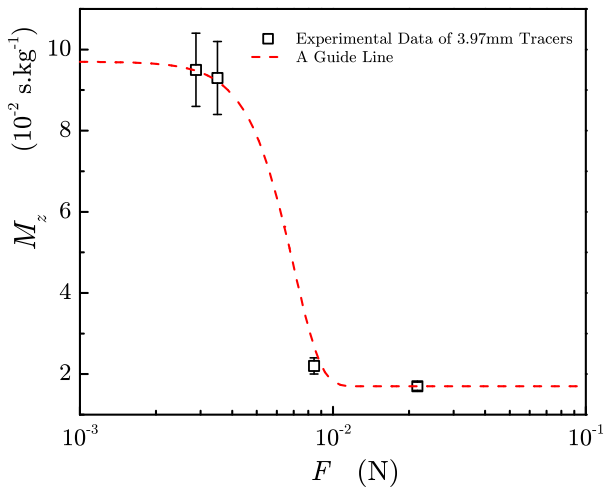}}}
\caption{(Color online) Mobility, $M_z$, versus the external force,
$F$, for $3.97$mm tracers in Packing 1. The black squares are
experimental data which is coming from different types
of tracer, they are (from left to right) nylon, delrin,
ceramic and brass. The red dashed guide line is a function
of $8.0 \exp{[-(x/0.007)^4]}+1.7$.} \label{M_F}
\end{figure}

In an effort to further test the concept of the effective
temperature as an intensive quantity, a linear response regime in
the system is of great interest. Such a regime would imply that
mobility and diffusivity are independent of the external
gravitational force as $F\to 0$. The external force is varied by
changing the density of the tracers of the same size. This is
realized experimentally in Packing 1 by the introduction of delrin
($\rho' = 1.36$) tracers of $3.97$mm diameter, the density of
which is higher than that of nylon ($\rho' = 1.12$).

Analysis of the trajectories reveals that the mobility is
approximately the same for both the delrin and nylon tracers with
the same diameter and is thereby independent of the external
force, as shown in Fig. \ref{D_M_T}b and Fig. \ref{M_F}.
Further, the external force
should have no effect on the diffusivity.  By calculating the
diffusivity of the non-tracer particles via dying acrylic tracers
and analyzing their trajectories, as shown in Fig. \ref{D_M_T}a,
the diffusion of the acrylic tracers of size $3.17$mm (for which
no external force is applied) is the same as the diffusion of the
delrin tracers of the same size (for which the gravitational force
is applied).  A further example of this property would be using
two different tracers with different sizes, but having the same
external force applied.  One would calculate two different values
of the diffusivity, due to the variation in tracer size, without
having any variation in external force.

Nonlinear effects appear for tracers heavier than delrin, implying
that mobility depends on the external force for large enough
forces. We find that for a $3.97$mm ceramic tracer $(\rho'=3.28)$
in Packing 1 the mobility is $M_{z\ \mbox{\scriptsize
ceramic}} = ( 2.2 \pm 0.2)
 \times 10^{-2} $ s/kg and for a brass tracer $(\rho'=8.4)$,
$M_{z\ \mbox{\scriptsize brass}} = ( 1.7 \pm 0.1)
 \times 10^{-2} $ s/kg as shown in Table \ref{table},
 smaller than the mobility of the
nylon and delrin tracers of the same size. This behavior is
expected since if a linear regime exists in the system, it will be
valid only within certain limits, i.e., $M_z$ remains a constant
for small value of external force, $F$,
as shown in Fig. \ref{M_F}. It is here that our experiments
approach the boundaries presented above for estimated of energy
scales for the sheared granular system.  Our effective temperature
measurements are therefore limited to those tracer particles for
which we experience a linear regime with respect to both mobility
and diffusivity.




Lastly, the experiment is again repeated for a different packing
of spherical particles, noted earlier as Packing 2. Having nearly
the same volume fraction of particles being both packings of
spherical particles, one would expect $T_{\mbox{\scriptsize eff}}$
to remain unchanged, as it is a measure of how dense the
particulate packing is (i.e. a large $T_{\mbox{\scriptsize eff}}$
implies a loose configuration, e.g. random loose packing, while a
reduced $T_{\mbox{\scriptsize eff}}$ implies a more compact
structure, e.g. random close packing). It is found that although
differences exist between the two packings with respect to
mobility and diffusivity, as shown in Fig. \ref{D_M_T}a,b, the
effective temperature remains approximately the same, as shown in
Fig. \ref{D_M_T}c. It should be noted that both packings are
composed of spherical particles and the statement regarding
effective temperature as a measure of particulate packing density
would not be true if the packings are composed of particles of,
for instance, different shapes, even if they have the same volume
fraction.

\begin{figure}
\centering{\resizebox{8cm}{!}{\includegraphics{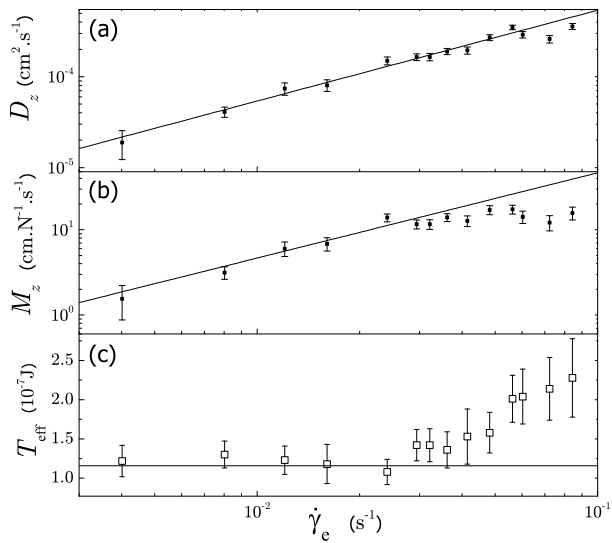}}}
\caption{The dependence of (a) diffusivity $D_z$, (b) mobility
$M_z$ and (c) effective temperature $T_{\mbox{\scriptsize eff}}$
on the shear rate $\dot\gamma_e$ for the $4.76$mm nylon tracers in
Packing 2. The solid lines in (a) and (b) are linear fitting [note
that the line in (b) is a fitting only for the first 6 data points
at the small value of shear rate $\dot\gamma_e$]. We find that
$D_z\sim\dot\gamma_e$ and $M_z\sim\dot\gamma_e$, while
$T_{\mbox{\scriptsize eff}}=D_z/M_z$ is approximately constant for
sufficiently small $\dot\gamma_e$. This quasi-static regime
coincides with the appearance of a rate-independent stress in
experiments \cite{tardos}, that $T_{\mbox{\scriptsize eff}}$ is
interpreted as the temperature of the jammed states. The height of
flat solid line in (c) is calculated from the slope of lines in
(a) and (b), which indicates a constant effective temperature
$T_{\mbox{\scriptsize eff}} = (1.2 \pm 0.2) \times 10^{-7}$J at
the small value of shear rate $\dot\gamma_e$.} \label{D_M_shear}
\end{figure}

\subsection{Shear Rate Dependence} \label{SRate}

We further explore the effective temperature as an intensive
quantity by analyzing diffusivity and mobility as a function of
the shear rate.  We show in Fig. \ref{D_M_shear} that the
effective temperature seems to become approximately constant, as
long as the particulate motion is slow enough such that the system
is very close to jamming. We find that

\begin{equation}
D_z\sim \dot\gamma_e, \ \ \ \ \dot\gamma_e \lesssim 0.06
\texttt{s}^{-1}
\end{equation}
\begin{equation}
M_z\sim \dot\gamma_e, \ \ \ \ \dot\gamma_e \lesssim 0.04
\texttt{s}^{-1}
\end{equation}
while $T_{\mbox{\scriptsize eff}}=D_z/M_z$ remains approximately
constant for sufficiently small $\dot\gamma_e$.

It is within this quasi-static range where the effective
temperature could be identified with exploration of the jammed
configurations. As it remains an important assumption of this
study that the system is being continuously jammed, shear rates
high enough to impact the effective temperature measurement imply
systems that are not continuously exploring jammed configurations.
As we study the nature of the jammed granular packings, it is
logical to presume that quasi-static shearing will provide systems
of interest. The limit of $T_{\mbox{\scriptsize eff}}$ as
$\dot\gamma_e\to 0$ may result in an effective temperature for the
static jammed configuration.  The quasi-static shear rate regime
observed could be analogous to the shear-rate independent regime
observed in the behavior of shear stress in slowly sheared
granular materials \cite{savage,tardos}. This solid friction-like
behavior has been previously studied \cite{savage,tardos} and
occurs when frictional forces and enduring contacts dominate the
dynamics. This regime has been also observed in recent computer
simulations of the effective temperature of sheared granular
materials \cite{ono,NingXu}. Our calculations of $T_{\mbox{\scriptsize
eff}}$ for systems close to jamming exclude the systems outside of
the quasi-static range, in accordance with prior studies.


\section{Outlook: Significance Of $T_{\mbox{\scriptsize eff}}$
 for a Statistical Mechanics of Grains}

In contrast to measurements of slow mode temperatures, exemplified
by $T_{\mbox{\scriptsize eff}}$, we also measure the temperature
of the fast modes as given by the root mean square (RMS)
fluctuations of the velocity of the particles. It should be noted
that these velocities are not instantaneous, as the time necessary
to obtain an instantaneous velocity is much smaller than the time
between measurements. Nevertheless, we can obtain an estimate of
the kinetic granular temperature, $T_{k}$, from $T_{k} =
\frac{2}{3}E_{k}$, where $E_{k} = \frac{1}{2}m\overline{v^{2}}$
with $\overline{v^{2}}$ the average kinetic energy of the grains.
We obtain $T_{k} = 9.17 \times 10^{10}k_{B}T$, or $3.77 \times
10^{-10}J$ and $T_{k} = 1.54 \times 10^{11} k_{B}T$, or $6.34
\times 10^{-10}J$, for $3.17$ mm and $3.97$ mm delrin tracers in
packing 1, respectively.  Here, $T = {298.15 K}$, the room
temperature, and $k_{B} = 1.3806504 \times 10^{-23}JK^{-1}$. This
kinetic granular temperature is smaller than $T_{\mbox{\scriptsize
eff}}$ and differs for each type of tracer indicating that it is
not governed by the same statistics. Similar results have been
obtained in experiments of vibrated granular gases \cite{menon}.
The significance of this result is that fast modes of relaxation
are governed by a different temperature. This result is analogous
to what is found in models of glasses and computer simulations of
molecular glasses (see for instance Refs.
\cite{ckp,liu1,mehta,coniglio,wolf}). In the glassy phase of these
models, the bath temperature is found to control the fast modes of
relaxation and a different, larger, effective temperature is found
to control the slow modes of relaxation. Similarly, we find a
granular bath temperature for the fast modes and a larger
effective temperature for slow modes of relaxation.

It is possible to identify $T_{\mbox{\scriptsize eff}}$ as the
property of the system governing the exploration of jammed
configurations.  As this particular non-equilibrium system remains
athermal, the 'bath' temperature in which the grains exist is
immaterial, as shown above. Particle diffusion is of the order of
several particle diameters over the time scale of the experiment
(see Fig. \ref{trajectory1} and \ref{D_M_T}a) implying that
exploration of the available jammed configurations occurs via
rearrangements of the particles outside their ``cages'',
suggesting that the trajectory of the system can be mapped onto
successive jammed configurations explored by the system.

Incorporating certain experimental conditions of reversibility,
and ergodicity, a statistical mechanics formulation may well
describe a jammed granular system \cite{edwards,mehta2}. Under the
primary assumption that different jammed configurations are taken
to have equal statistical weight, observables can be calculated by
``flat'' averages over the jammed configurational space
\cite{edwards,kurchan,bklm,coniglio2,mk,luck,mehta3}. This
assumption, advocated by Edwards and collaborators, has been
thoroughly debated in the literature (see for instance
\cite{coniglio,cavendish}). Existing work suggests the effective
temperature obtained by applying a fluctuation-dissipation theory
to non-equilibrium systems is analogous to performing a ``flat''
average over the jammed configurational space, at least for
frictionless systems \cite{mk}. Additionally, the effective
temperature can be identified with the compactivity introduced in
\cite{edwards}, resulting from entropic calculations of the
granular packing \cite{kurchan,bklm,mk}. Experimentally testing
these ideas is difficult as the entropy of the jammed
configurations is not easily measured, and it is not possible to
obtain the compactivity from entropic considerations in the
present study.

The exploration of reversible jammed states in granular matter
bears similarity to that of inherent structures in glasses.
Inherent structures form a network of attractive basins within an
energy landscape, and the system explores these basins as governed
by their stability over the slow-relaxation time of the glass. It
should be noted, however, that there exists a crucial difference
between glasses and grains.  In liquids energy remains conserved,
while energy is dissipated in granular systems through frictional
contact and path dependent forces between grains. Thus, a driven
granular system will quickly come to a mechanically stable, or
jammed, state after the removal of the driving forces. By its
nature, energy is not conserved in a granular system. As energy
conservation is the crucial property used to define an energy
ensemble in statistical mechanics, the use of energy to
characterize granular systems is questionable.  Thus, while
$T_{\mbox{\scriptsize eff}}$ seems to imply the exploration of
reversible jammed states within an energy ensemble, with

\begin{equation}
P(E)\sim\text{e}^{-\frac{E}{T_{\mbox{\scriptsize eff}}}}
\end{equation}
describing the nature of the exploration, the validity of the
energy ensemble to describe granular matter in the absence of
energy conservation remains an open question.  Here, we set the
analogous Boltzmann constant for grains equal to unity for
simplicity.

Noting the drastic difference between the bath temperature and the
effective temperature in a granular system, we are inspired
towards a more careful analysis of the energy ensemble in slowly
driven granular systems.  The work of Edwards has promoted the
concept of a volume ensemble, where the free volume per grain in a
static granular system replaces the energy as the conserved
quantity of the non-equilibrium system, at a particular volume
fraction \cite{edwards,mehta2}.  The basis for using the volume
ensemble stems from the ability to conserve volume in a given
packing and additivity of volume per grain.  Further, it is
possible to explore the configuration of states at a fixed volume,
via experiment or simulation.  The statistical mechanics is then
derived using methods similar to Boltzmann statistics for
equilibrium systems. From these methods, one can obtain the
compactivity, $X$, as a derivative of the entropy with respect to
the volume, enabling the calculation of an equation of state in
the volume ensemble as follows.

\begin{equation}
X^{-1} = \frac{\partial{S}}{\partial{V}} \label{compactivity}
\end{equation}

The compactivity, $X$, is thereby assumed to be an equilibrium
measure of a system within the framework of the volume ensemble,
much like the bath temperature of the energy ensemble.  This assumption can be realized by performing an ABC experiment and testing a zero-th law of thermodynamics for volumes \cite{cavendish}. According to the zero-th law of thermodynamics, if system A and C are in thermal equilibrium with system B respectively, then A and C are in thermal equilibrium with equal temperature. In granular system, such an experiment would require two granular systems with distinct volumes, $V_1$ and $V_2$, with the same $X$. Bringing these two systems together should result in a granular system of volume $V = V_1 + V_2$, at the same $X$, if the assumption is valid.
This experiment is feasible due to the fact that it is always
possible to prepare a system at a given volume fraction and will
be the subject of future study and experiment, facilitated by
recent theoretical findings \cite{chaoming_nature}.

Similar to the conservation of volume, boundary stress may also be a
conserved quantity in jammed granular systems, and Edwards
statistical mechanics for volume distributions could be applied
analogously to the distribution of boundary stresses, $\Pi$, or forces,
referred to as the force ensemble \cite{Edwards_new,Henkes}.  The
angoricity, $A$, is calculated as the derivative of the entropy
with respect to the boundary stress, and an additional equation of state is
thereby achieved as follows:

\begin{equation}
A^{-1} = \frac{\partial{S}}{\partial{\Pi}}
\label{angoricity}
\end{equation}

This result can be combined with that of the volume ensemble in an
effort to accurately define the statistical mechanics of static
jammed granular matter.  Such an approach remains a topic of
ongoing research.

However, slowly driven granular systems introduce yet another
ensemble, the energy ensemble, from which the above defined
$T_{\mbox{\scriptsize eff}}$ is derived.  While the above results
reveal that $T_{\mbox{\scriptsize eff}}$ does not tend to zero as
the magnitude of the driving force decreases, indicating
extrapolation to a non-zero static quantity, it remains unclear
how the effective temperature may relate to the compactivity and
angoricity as defined by Edwards statistics. Are we defining a new
static quantity by determining the static limit of
$T_{\mbox{\scriptsize eff}}$, or are we expanding the statistical
mechanics of jammed granular matter to include dynamic systems by
relating $T_{\mbox{\scriptsize eff}}$, $X$ and $A$?
$T_{\mbox{\scriptsize eff}}$ is obtained in the quasi-static limit
$\dot{\gamma_{e}}\rightarrow 0^{+}$, while the volume and force
ensembles correspond to $\dot{\gamma_{e}} =  0$, exactly.  Is it
possible that a relation between $T_{\mbox{\scriptsize eff}}$, $X$
and $A$ can be expected?

There exists the further requirement of energy conservation for
the validity of a Boltzmann approach that would guarantee:

\begin{equation}
T_{\mbox{\scriptsize eff}}^{-1} = \frac{\partial{S}}{\partial{E}}
\label{effective}
\end{equation}

As discussed above, energy is constantly dissipated in a driven
granular system, through Coulomb friction and path-dependent
tangential forces between grains.  However, the input of energy by
the external driving force brings the system to a steady state
where the average energy is constant over the time-scale of the
experiment. This steady state energy could be likened to the
conserved variable in a statistical formalism depicted in Eq.
\ref{effective}, thereby introducing a thermodynamic meaning for
$T_{\mbox{\scriptsize eff}}$.

In a compressed emulsion system the absence of Coulomb friction
and inter-particle tangential forces greatly simplifies the
formalism \cite{brujic}.  Jamming occurs due to osmotic
pressure, and the system remains athermal as a result of the large
particle size. A well defined potential energy exists due to the
absence of tangential forces, corresponding to the deformation of
the particles at the inter-particle contact points.  Therefore, a
restriction to use the energy ensemble in an effort to describe a
jammed system is lifted, as frictional tangential forces no longer
hinder energy conservation.

Computer simulations of frictionless emulsion droplets
\cite{cavendish} incorporate a simulated annealing method
employing an auxiliary temperature to sample the available jammed
configurations.  The simulated annealing method assumes a
Boltzmann distribution, or a flat average assumption, over the
jammed states of the emulsion. The $T_{\mbox{\scriptsize eff}}$
obtained by Eq. \ref{effective} with simulated annealing methods
\cite {mk} is very close in value to the $T_{\mbox{\scriptsize
eff}}$ obtained via the FDT calculations as in the present work.
Such a result could indicate that ergodicity holds in this
frictionless system, a further justification of the methods
presented herein. Therefore, a firmer basis for the validity of
using the $T_{\mbox{\scriptsize eff}}$ obtained in the
quasi-static limit to describe the statistical mechanics of the
same system at the static limit is achieved.

Further, $T_{\mbox{\scriptsize eff}}$ remains approximately
constant with varying tracer particle size, implying a zero-th law
of thermodynamics for slowly sheared jammed granular systems.
These statements further provoke the necessity for an ABC
experiment, to test the zero-th law for the effective temperature,
as well as similar experiments for the compacitivity and
angoricity.  Such experiments may enlighten us to understand under
what conditions $P(E)\sim\text{e}^{-\frac{E}{T_{\mbox{\scriptsize
eff}}}}$, $P(V)\sim\text{e}^{-\frac{V}{X}}$ and
$P(\Pi)\sim\text{e}^{-\frac{\Pi}{A}}$ may be valid in describing
the statistics of the jammed and nearly jammed granular systems.

At this point, we believe that the most prominent direction is the
exploration of the volume and pressure ensembles.  Our
understanding is that these ensembles may be sufficient to
characterize the jammed state of granular matter, while the energy
ensemble may be necessary for slowly moving granular
systems.  These are open questions at the present time.  Recent
papers in the theory and simulation front suggest that the
compactivity characterizes the system into a phase diagram at the
isostatic point, while the angoricity will be necessary to
describe the pressure ensemble of compressible granular matter
\cite{chaoming_nature}.

\section{Summary}

In summary, this study focuses on the dynamics of slowly sheared
granular matter in a 3D Couette cell.  A mixture of spherical,
transparent and bi-disperse grains are confined between two
cylinders, having walls roughened by glued identical grains, with
the inner cylinder rotated via motor. We compact the grains by
means of an external pressure in the negative z-direction.  Fluid
matching the density and refractive index of the grains partially
fills the cell, allowing tracking of tracer particle trajectories
as a function of time. Tracers of varying density and size are
used. Multiple cameras track the tracer particle positions
relative to the cylinders.

We find that the angular velocity of the tracer particles,
$\omega_\theta(r)$, follows an exponential relation with $r$,
defined by the type of packing and geometry of the Couette cell.
The velocity of the last
layer of grains is non-zero, such that the shear band is located
at the outer cylinder and ensures no formation of shear bands in
the bulk. Near the outer cylinder $\omega_\theta(r)$ decays slowly
with increasing $r$,
such that
$\omega_{\theta}(r)$ can be approximated linearly with a constant
local shear rate. The constant local shear rate ensures that the
mobility and diffusivity of tracers, dependent on local shear
rate, remain approximately independent of $r$. We define this
region the ``constant mobility and diffusivity region'', or the
CMD region.

An ``effective temperature'',
$T_{\mbox{\scriptsize eff}}$, is realized by a
fluctuation-dissipation relation generalized to granular
materials.
Statistical measurements are confined
exclusively to the CMD region.
The mobility in the vertical direction, $M_z$,
is found to be proportional to the shear rate, $\dot\gamma_e$, for
small enough values of $\dot\gamma_e$.
As $D_z$ is also found proportional
to shear rate, collapsing all the $D_z(r)$ for various shear rates
shows a plateau in the CMD region.  An approximately constant
effective temperature is obtained from measurements of the
mobility and diffusivity, under a constant external applied force,
and with sufficiently small shear rates. This effective
temperature is calculated by an analogous equation used in
equilibrium statistical mechanics. We find this effective
temperature to be independent of the tracer particle properties,
and dependent only on the packing density of the system.  While
this result describes an intensive property of the system, it
remains an important future study to test the effective
temperature against the laws of thermodynamics. More specifically,
a test of the zeroth-law of thermodynamics with respect to these
non-equilibrium jammed systems could expand the scope of
$T_{\mbox{\scriptsize eff}}$ beyond that of an intensive quantity
of a particular system.  A well defined effective temperature in
the radial direction may exist, though its existence would require
a constant external force applied in the radial direction.

The probability distribution of the displacements in the radial
direction, $P(\Delta r)$, reveals exponential fluctuations.
The analysis of the fluctuations
reveals a power law, sub-diffusive, process, $\langle\Delta
r^2\rangle\sim{\Delta t}^\alpha$, with $\alpha$ less than unity. A
similar analysis for fluctuations in the angular direction reveal
a super-diffusive process,
$\langle\Delta\theta^2\rangle\sim{\Delta t}^\beta$, with $\beta$
greater than unity. Lastly, the probability distribution of the
displacements in the vertical direction are found have a Gaussian
distribution such that $\langle\Delta z^2\rangle\sim{\Delta t}$.
It is this linearity that defines vertical displacement as a
diffusive process, and allows for the use of the
Fluctuation-Dissipation relation to calculate the diffusivity in
the vertical direction.  We further discover a linear relationship
between angular displacement and the time between measurements
$\Delta{\theta_e}\sim{\Delta t}$, such that all mean square
fluctuations can be defined in terms of $\Delta{\theta_e}$ for the
small shear rates of our experiments.

In the CMD region, the linear approximation of $\omega_\theta(r)$
proportional to approximately constant external shear rate,
$\dot\gamma_e$, allows for the collapsing of all tracer particle
velocity curves via dividing $\omega_\theta(r)$ by the shear rate.
This collapse reveals a periodic shape with a small amplitude and
periodic length roughly equal to the grain size. The effect is
shown to be weaker in packings with smaller size grains. We
further apply this remarkable scaling feature $v_z(r)$ and
$v_r(r)$, achieving similar results.

It is important to note that the effective temperature, defined in
this study for small shear rates, does not remain constant as the
shear rate increases.  While previous studies have discovered an
an increasing effective temperature via simulations, we have
measured diffusivity and mobility separately in an effort to
calculate $T_{\mbox{\scriptsize eff}}$ through a
fluctuation-dissipation relation. We find the diffusivity in the
z-direction remains approximately constant throughout the range of
shear rates used in this experiment, while the mobility in the
z-direction approaches a plateau.
exclusively increasing $T_{\mbox{\scriptsize
eff}}$.
Such an effect in the radial direction would be of great interest for
future studies in sheared granular dynamics.

The nearly constant value of $T_{\mbox{\scriptsize eff}}$ with
respect to varying tracer particle size indicates that a zero-th
law of thermodynamics for slowly sheared jammed granular systems
could be valid and prompts one to perform an ABC experiment, fully
testing the zero-th law for the effective temperature.


As we work towards a more complete description of the statistical
mechanics of jammed granular matter, we strive to incorporate the
varied statistical ensembles into one fundamental picture. These
ensembles include the energy ensemble, as described herein, along
with the volume and force ensembles, as proposed by Edwards.  Such
an incorporation may link static quantities of compacitivity and
angoricity, describing volume and force ensembles, respectively,
to the dynamic effective temperature presented in this study,
derived from the energy ensemble.  The exact nature of the
relation between such quantities remains an open topic.
Ultimately, these quantities will help to develop a thorough
statistical description for jammed granular matter and reveal an
equation of state.   A deeper topic of concern is the formation of
a clear definition of energy in jammed granular matter.  Energy is
not conserved in frictional systems and it remains open to debate
as to how one would incorporate energy into the statistical
mechanics.

One possible approach to describe the energy of jammed systems is
to consider the similarities between the inherent structure
formalism of glasses and the exploration of jammed states in
granular matter, at least for the case of frictionless granular
systems. Inherent structures probe a network of potential energy
basins within an energy landscape. Such an approach toward the
jammed states of granular matter may assist in understanding
exactly what is meant by energy within the framework of a
non-equilibrium system.

\vspace{1cm}

We are deeply grateful to M. Shattuck for help in the design of
the Couette cell and J. Kurchan and  J. Bruji\'c for discussions.
We acknowledge financial support from the DOE, and NSF.

\clearpage

\end{document}